\documentclass[12pt,a4paper,oneside]{article}

\usepackage{graphicx,amssymb,amsmath}
\usepackage{setspace} 
\usepackage[linkcolor={blue},citecolor={blue},colorlinks=true,linktocpage,urlcolor={blue}]
{hyperref}
\usepackage{cite}
\usepackage[margin=2cm]{geometry}
\usepackage[dvipsnames]{xcolor}
\definecolor{darkblue}{rgb}{0,0,0.54}
\usepackage{tikz}
\usetikzlibrary{patterns}
\usetikzlibrary{calc}
\usepackage{multirow} 
\usepackage{appendix}
\usepackage{bbm}
\usepackage{import}
\usepackage{tensor}
\usepackage{pdfpages}
\usepackage{cancel} 
\usepackage{slashed} 
\usepackage{xparse} 
\usepackage{tabu} 
\usepackage{adjustbox} 
\usepackage{simpler-wick}
\usepackage[bb=boondox]{mathalfa}
\usepackage{dsfont}
\usepackage{subcaption}

\numberwithin{equation}{section} 
\usepackage{empheq} 

\graphicspath{{Figures/}} 

\newcommand{\pder}[2]{\frac{\partial#1}{\partial#2}} 

\DeclareDocumentCommand \at { o m }
{
           \IfNoValueTF {#1}
             {\big |_{#2}}
             {\left.#1\right|_{#2}}
}

\usepackage{titling} 
\usepackage[affil-it]{authblk} 

\usepackage{tocloft} 
\title{Dualities between fermionic theories and the Potts model}
\author{Vladimir Narovlansky}

\affil{\small Princeton Center for Theoretical Science and Physics Department,
Princeton University, Princeton, NJ 08544, USA}
\date{\small \texttt{narovlansky@princeton.edu}}

\begin{document}

\begin{titlingpage}
    \maketitle
    \begin{abstract}
    We show that a large class of fermionic theories are dual to a $q \to 0$ limit of the Potts model in the presence of a magnetic field. These can be described using a statistical model of random forests on a graph, generalizing the (unrooted) random forest description of the Potts model with only nearest neighbor interactions. We then apply this to find a statistical description of a recently introduced family of $OSp(1|2M)$ invariant field theories that provide a UV completion to sigma models with the same symmetry.
    \end{abstract}
\end{titlingpage}

\tableofcontents

\section{Introduction}

The Potts model \cite{Potts:1951rk,Wu:1982ra} is an immediate generalization of the Ising model, where at every site of a lattice there is a degree of freedom that can obtain $q$ possible states, with $q=2$ being precisely the Ising model. Nearest neighbors interact among themselves according to their state. For example, whenever they are in the same state, the energy of the system gets a particular contribution, analogously to the case of spins of particles that point in the same direction.

Given this arguably simplest statistical system one could imagine, it is natural to wonder how rich its statistical behavior is. That is, it is interesting to know how many different phases can such a simple model capture. In this work, we would like to show that a wide class of fermionic theories with general interactions are dual to the $q \to 0$ limit of the Potts model in the presence of a magnetic field.\footnote{We should point out that we use `fermions' to refer to anticommuting degrees of freedom throughout. These are not necessarily fermions in the usual sense.} This is done in section \ref{sec:duality}, and the precise relation is given in Eq.\ \eqref{eq:equiv_ferminic_Potts}.

The $q \to 0$ limit of the Potts model with only nearest neighbor interactions is known to be equivalent to a statistical model of unrooted \emph{spanning forests} on the graph defined by the lattice on which the Potts degrees of freedom reside. We use a generalization of the spanning forests theory in order to argue for the duality above. We do this by relating it to fermionic theories similarly to \cite{Caracciolo:2004hz} on the one hand, and to the Potts model generalizing the idea of Fortuin and Kasteleyn \cite{kasteleyn1969phase,Fortuin:1971dw} on the other hand.

We then apply this equivalence to find a statistical description of a class of $OSp(1|2M)$ invariant theories that were recently introduced \cite{Klebanov:2021sos} in terms of the Potts model as well as spanning forests (section \ref{sec:fermionic_OSp}).
In Ref.\ \cite{Klebanov:2021sos}, field theories with $OSp(1|2M)$ symmetry were constructed and were suggested to be the UV completions of sigma models with the same symmetry. The case $M=1$ was studied in \cite{Fei:2015kta}.
The $q \to 0$ limit of the Potts model with nearest neighbor interactions, or equivalently the statistical theory of (unrooted) random forests, are known to correspond to the sigma model with $OSp(1|2)$ symmetry \cite{Caracciolo:2004hz,Jacobsen:2003qp}. The field theory with $M=1$, which is suggested to provide a UV completion of this sigma model, was observed in \cite{Fei:2015kta} to have critical exponents matching to those of spanning forests. Here, we would like to understand this relation and extend it to general $M$, with particular emphasis on $M=2$.
The $M=1$ field theory is cubic in the fields and becomes weakly coupled in six dimensions. This suggests that the upper critical dimension for unrooted spanning forests is six. In fact, numerical evidence for this was given in \cite{Deng:2006ur} (see also \cite{BenAliZinati:2017vjy}).
In a general spanning forest representation, as we describe below, the roots of the different trees can be correlated. We show in section \ref{sec:M_2_forest} how this can be eliminated, and demonstrate this by analyzing spanning forest descriptions of the $M=2$ theory.

An additional motivation to study such vector theories with anticommuting scalars comes from the de Sitter / Conformal Field Theory (dS/CFT) correspondence. In dS/CFT, gravity in dS space, having a positive cosmological constant, is related to a non-unitary CFT living in Euclidean space \cite{Strominger:2001pn}.  This is motivated by the asymptotic symmetries of dS gravity, with appropriate boundary conditions obtained by analytic continuation of those in AdS \cite{Brown:1986nw}.
It was proposed \cite{Anninos:2011ui} that the dual to the minimal higher spin theory in 4-dimensional de Sitter is a theory of anticommuting scalars with quartic interactions \cite{LeClair:2006kb,LeClair:2007iy}, having $Sp(N)$ symmetry. The idea is that higher spin theory in dS is obtained by the one in AdS by continuing the cosmological constant $\Lambda $ to $-\Lambda $. Using the relation $N \sim \frac{1}{G_N\Lambda } $ of the boundary central charge to Newton's constant $G_N$, this means taking $N \to -N$. The transformation $N \to -N$ is implemented by turning the commuting variables into anticommuting. This is seen using a Hubbard–Stratonovich field since each anticommuting loop gives a minus sign and comes with a factor of $N$. Therefore, going from bosons where each loop is assigned a value of $N$ to fermions implements taking $N \to -N$. We give an alternative non-perturbative argument for this claim in appendix \ref{sec:NToMinusN}.

A UV completion of the quartic vector theory in dimension beyond four dimensions is the theory with $M=1$ potential that we discuss in section \ref{sec:fermionic_OSp}, with an arbitrary number of anticommuting fields. This was suggested in \cite{Fei:2015kta} to provide a higher spin theory in dS in higher dimension.

\section{Duality between fermionic theories and interacting Potts models} \label{sec:duality}

Consider a many-body system (or a regularized field theory) that we describe using an undirected graph $G$ with vertices $V$ and edges $E$. Let the graph be edge-weighted so that to every edge $e$ we assign a weight $w_e$. We can assume that every two vertices are connected by at most one edge.\footnote{This is so, because only the sum of weights of edges connecting two vertices enters in the Laplacian matrix below. Note also that there are no edges connecting a vertex to itself.} We can also denote by $w_{ij} =w_{ji}$ the weight of the edge connecting vertices $i$ and $j$ in case they are connected by an edge, and zero otherwise. The Laplacian matrix of $G$ is a symmetric $|V| \times |V|$ matrix with entries $L_{ij} =-w_{ij} $ for $i \neq j$ and $L_{ii} =\sum _{k \neq i} w_{ik} $, so that each row and column sum to zero. For example, the graph $G$ can be a bounded cubic lattice with nearest neighbors connected by edges. For unit weights, the Laplacian matrix is simply the discretized Laplacian.

A rather general fermionic theory is defined by providing a lattice $G$ and placing Grassmann variables $\theta _i$ and $\bar \theta _i$ on each vertex $i \in V$. The interactions we consider are specified by subgraphs of $G$ labeled by $\Gamma $ with vertex sets $V_{\Gamma } $, so that the partition function is given by
\begin{equation} \label{eq:fermionic_theory_no_edges}
\int D\theta D\bar \theta \, e^{\bar \theta L\theta +\sum _{\Gamma } t_{\Gamma } \prod _{i \in V_{\Gamma } } \left( \bar \theta _i \theta _i\right) } 
\end{equation}
where $t_{\Gamma } $ are the coupling constants.
As explained below, without loss of generality, we will take the $\Gamma $'s to have no edges for the purpose of this formula.

We would like to show that this theory is equivalent to the Potts model with interactions beyond nearest neighbors. The Potts model can be thought of simply as a generalization of the Ising model, where instead of two states, the degrees of freedom have $q$ states. In terms of the graph description above, we place variables $\sigma _i$ on every vertex $i$, taking values $\sigma _i \in \{1,\cdots ,q\}$. Similarly to the Ising model, we can also introduce a fixed external magnetic field with direction $h \in \{1,\cdots ,q\}$. Nearest neighbor interactions are given simply by Kronecker delta symbols $\delta _{\sigma _i,\sigma _j} $, where we consider $i$ and $j$ to be nearest neighbors if they are connected by an edge in the graph. We will introduce higher interactions by coupling not only nearest neighbor pairs, but also next-to-nearest neighbor terms and beyond (if necessary) with Kronecker delta symbols.\footnote{This can be thought of as placing the Potts model on a hypergraph rather than a graph.} We will find it useful to couple them to the external magnetic field.

Specifically, the claim is that the fermionic theory \eqref{eq:fermionic_theory_no_edges} is equivalent to a $q \to 0$ limit of a Potts model, such that
\begin{equation} \label{eq:equiv_ferminic_Potts}
\begin{split}
& \int D\theta D\bar \theta \, e^{\bar \theta L\theta +\sum _{\Gamma } t_{\Gamma } \prod _{i \in V_{\Gamma } } \left( \bar \theta _i \theta _i\right) } = \lim _{q \to 0} q^{-|V|/2} \sum _{\sigma _i \in \{1,\cdots ,q\}} e^{-H} .
\end{split}
\end{equation}
The Hamiltonian of the theory is
\begin{equation} \label{eq:equiv_ferminic_Potts_Hamiltonian}
H = -\sum _{e=\langle ij\rangle } \log (1+\sqrt{q}w_e) \delta _{\sigma _i,\sigma _j} - \sum _{\Gamma } \log \left( 1+t_{\Gamma } q^{|V_{\Gamma }| /2} \right) \prod _{i \in V_{\Gamma } } \delta _{\sigma _i,h}
\end{equation}
where in the first sum we go over nearest neighbors.\footnote{Note that we can replace each of the three explicit occurrences of $\sqrt{q}$ in the formula by $q^{\alpha } $ for any $0<\alpha <1$ and the formula will still be valid.}

In the remainder of this section, we prove \eqref{eq:equiv_ferminic_Potts}. There are two steps: first, we express the fermionic theory using rooted spanning forests, and then we extend the idea of Fortuin and Kasteleyn \cite{kasteleyn1969phase,Fortuin:1971dw,Edwards:1988ba,Jacobsen:2003qp} in order to relate the forest description to the Potts model.

\begin{figure}[h]
\centering
\includegraphics[width=0.4\textwidth]{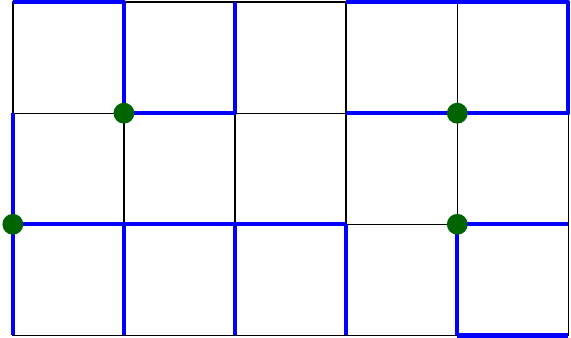}
\caption{In this example, the graph is a 2-dimensional lattice. We need to go over all the possible spanning forests, with one such forest shown in heavy blue lines. For an interaction $t \sum _i \bar \theta _i\theta _i$ we mark the trees with subgraphs consisting of a single vertex. Therefore, the forests are rooted, where each root is shown by a heavy green dot. We need to go over all the possible roots, and each root is assigned a factor of the coupling constant $t$.}
\label{fig:example_spanning_rooted_forests}
\end{figure}

To explain the spanning forest description more broadly, let us allow for the moment the subgraphs $\Gamma $ to have edges $E_{\Gamma } $, and following \cite{Caracciolo:2004hz} define for each such subgraph
\begin{equation}
Q_{\Gamma } = \left( \prod _{e \in E_{\Gamma } } w_e\right)  \left( \prod _{i \in V_{\Gamma } } \bar \theta _i \theta _i \right) .
\end{equation}
The fermionic theory with interactions given by the $\Gamma $'s can be expressed using a statistical model of spanning subgraphs $H$ of $G$, as was shown in \cite{Caracciolo:2004hz}. (Recall that a spanning subgraph of $G$ is a subgraph having the same vertex set as that of $G$.) We will need some generalization of the description in \cite{Caracciolo:2004hz}, in which we allow the $\Gamma $ subgraphs to be not necessarily connected. Following a derivation similarly to \cite{Caracciolo:2004hz} we get the description
\begin{equation} \label{eq:general_spanning_forests_formula}
\begin{split}
& \int D\theta  D\bar \theta \,  e^{\bar \theta  L \theta  + \sum _{\Gamma } t_{\Gamma } Q_{\Gamma } } = \sum _{\substack{H=(H_1,\cdots ,H_l) \text{ spanning} \\ H_i \text{ disjoint}}} \left( \prod _{i=1} ^l W(H_i)\right)  \prod _{e \in H} w_e .
\end{split}
\end{equation}
We now explain this formula. In the RHS, we should go over all the spanning subgraphs $H$ of $G$ and the ways to decompose them into disjoint components $H_i$. It is important that the $H_i$ are \emph{not necessarily connected}. Each subgraph $H$ is first assigned the value of the graph, which is the product of the weights of its edges. The function $W(H_i)$ for a subgraph $H_i$ is defined by
\begin{equation}
W(H_i)= \sum _{\Gamma \prec H_i} t_{\Gamma } 
\end{equation}
where we sum over all the markings of $H_i$. A subgraph $\Gamma $ from the set of interactions marks $H_i$, which is denoted by $\Gamma \prec H_i$, if they have the same number of connected components and each connected component of $H_i$ contains exactly one connected component of $\Gamma $, and in addition any cycle in $H_i$ comes entirely from a cycle in $\Gamma $.
In other words, for every subgraph $H$, the formula instructs us to \emph{go over all the ways to mark the different components using the interactions}.
If $\Gamma $ contain no cycles, the cycles condition above means that we only need to sum over spanning forests $H$. (A forest is a graph containing no cycles, while a tree is a connected graph with no cycles.)
In that case, and in particular if $\Gamma $ consist of only vertices, we obtain a statistical model of rooted forests, with the marks providing the roots.
Note that in the case that the $\Gamma _i$ are connected, we only need to consider connected $H_i$, and this is the description in \cite{Caracciolo:2004hz}. An example is shown in Fig.\ \ref{fig:example_spanning_rooted_forests}.

\begin{figure}[h]
\centering
\includegraphics[width=0.5\textwidth]{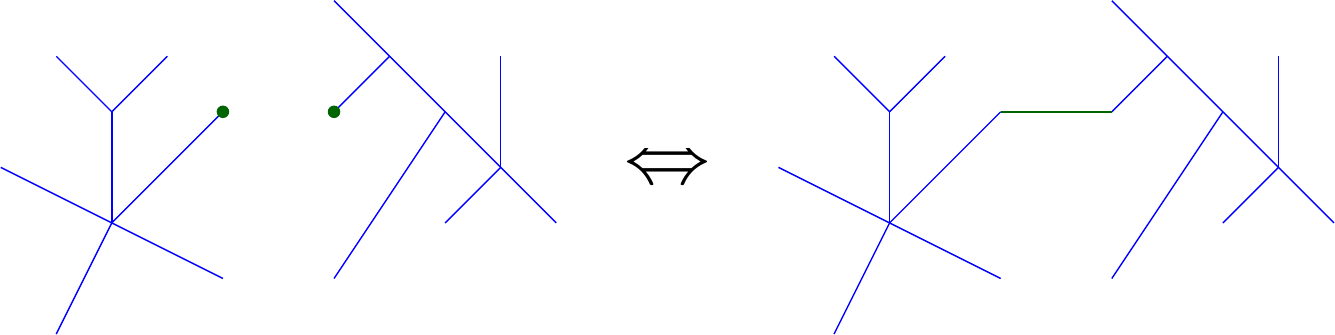}
\caption{Equivalence of a disconnected marking of a pair of trees with a connected marking of a single tree. The tree on the right hand side has an additional edge, and it is marked by a coupling constant.}
\label{fig:trade_nn_marking}
\end{figure}

Note that the difference between including edges in $\Gamma $ or not, is only whether we include the constants associated to the edges or we do not. Since we can absorb these in the couplings $t_{\Gamma } $, we may as well restrict to interactions without edges, and this is what we have done in \eqref{eq:fermionic_theory_no_edges}.
In addition, this provides a consistency check on the formula \eqref{eq:general_spanning_forests_formula}.
Namely, we can compare for example a subgraph consisting of two vertices and an edge $Q_{\Gamma } =w_{ij} \bar \theta _i \theta _i \bar \theta _j \theta _j$ (with $i,j$ nearest neighbors) to $Q_{\Gamma '} =\bar \theta _i \theta _i \bar \theta _j \theta _j$ without the edge. While the former corresponds to a connected subgraph, the latter is disconnected. However, they only differ by a constant and should give the same result for appropriate couplings. In the formula \eqref{eq:general_spanning_forests_formula} we could include either term, and the result is the same. Indeed, consider an $H_i$ with two connected components, such that $i$ and $j$ appear in each one, as on the left of Fig.\ \ref{fig:trade_nn_marking}. In this case, the former interaction does not mark this $H_i$ while the latter does. On the other hand, considering the same $H_i$ with the edge $ij$ added, we get a connected subgraph. Now the former interaction marks this $H_i$, as on the right of Fig.\ \ref{fig:trade_nn_marking}, while the latter does not. With the appropriate couplings, we get the same result, and we find a correspondence between subgraphs marked by the first interaction and other subgraphs marked by the second one. Since we sum over all subgraphs, the result is the same.

Going back to the form \eqref{eq:fermionic_theory_no_edges}, the interactions include no cycles, and hence the spanning subgraph statistical model becomes a spanning forests description.
We now prove \eqref{eq:equiv_ferminic_Potts} using this description. Using the identity $1+c \delta (\cdots )=e^{log(1+c) \delta (\cdots )} $, we can write the Potts side of the duality as
\begin{equation}
\lim _{q \to 0} q^{-|V|/2} \sum _{\sigma _i} \prod _{e=\langle ij\rangle } \left( 1+\sqrt{q}w_e \delta _{\sigma _i,\sigma _j} \right)  \cdot \prod _{\Gamma } \left( 1+t_{\Gamma } q^{|V_{\Gamma } |/2} \prod _{i \in V_{\Gamma } } \delta _{\sigma _i,h} \right) .
\end{equation}
Let us construct a subgraph of $G$ with edges $A$ defined as follows. An edge $e$ belongs to $A$ if in the first product we choose the Kronecker delta term when performing the product. This gives a subgraph $H$ of $G$; we will take all the vertices to belong to $H$ as well by definition, so that $H$ is spanning. We thus have a sum over spanning subgraphs of $G$. We interpret similarly the second product as generating markings, with the $t_{\Gamma } $ term chosen if we use the $\Gamma $ marking.

Each edge in $A$ is given a factor of $\sqrt{q}w_e$, and moreover we have a factor of $q$ for every (connected) component of $H$ from the Kronecker delta terms and the sum over the spins. If, however, such a connected component is marked once, it scales instead of $q$ as $q^{1/2} $ and so is more dominant. If it is marked twice or more, it scales at least as $q$ and so is suppressed. We see that if we fix $A$, the dominant contribution comes when every connected component is marked exactly once. In this dominant contribution, such an $A$ is assigned the value
\begin{equation}
\sum _{\text{markings }M} \, \prod _{\Gamma \in M } t_{\Gamma } \prod _{e \in A} w_e \cdot q^{\frac{1}{2} (|A| + k(H) )}
\end{equation}
where $k(H)$ is the number of connected components in $H$ and $|A|$ is the number of edges in $H$. For any graph, $|A|+k(H) \ge |V|$ and equality is attained for forests. We see that the dominant contribution goes as $q^{|V|/2} $ and is saturated by forests $H$. The dominant contribution is then $q^{|V|/2} \sum _{\substack{\text{spanning}\\\text{forests }H}} \sum _{\text{markings }M} \, \prod _{\Gamma \in M} t_{\Gamma } \prod _{e \in A} w_e$. We recover precisely the forest description \eqref{eq:general_spanning_forests_formula}.

\section{Theories with $OSp(1|2M)$ symmetry} \label{sec:fermionic_OSp}

Consider the sigma model with fermionic hyperbolic target space $\mathbb{H}^{0|2M} $ for a given $M \ge 1$. It has the following Euclidean action for a commuting field $\sigma $ and anticommuting $\theta ^i$
\begin{equation} \label{eq:OSp_sigma_model}
S=\frac{1}{2g^2} \int d^d x \left( (\partial _{\mu } \sigma )^2+2 \partial _{\mu } \bar \theta ^i \partial ^{\mu } \theta ^i \right) 
\end{equation}
where summation over $i=1, \cdots ,M$ is implicit, with the constraint
\begin{equation} \label{eq:OSp_sigma_model_constraint}
\sigma =\left( 1-2 \bar \theta ^i \theta ^i\right) ^{1/2} .
\end{equation}

This sigma model has an $OSp(1|2M)$ supergroup symmetry.
As a brief reminder, recall that $OSp(1|2M)$ is the Lie supergroup of linear transformations on one bosonic variable and $2M$ fermionic variables preserving the bilinear form $\sigma^2+2 \bar \theta ^i \theta^i$. The transformation mapping bosonic to bosonic variables and fermionic to fermionic variables constitute the even grade part of the group, while those mixing variables of different statistics are the odd grade elements corresponding to anticommuting generators.

Note also that the constraint $\sigma ^2+2\bar \theta ^i \theta ^i=1$ has two solutions, while we restrict to the single solution above. This can be interpreted either as a superhemisphere in $\mathbb{S} ^{0|2M} $ or as a hyperbolic target superspace $\mathbb{H} ^{0|2M} $ \cite{Bauerschmidt:2019mhu,Bauerschmidt:2021nfm} (by multiplying the defining equation by $(-1)$). Including both solutions to the constraint can be implemented by incorporating an additional Ising degree of freedom as was analyzed in \cite{Jacobsen:2005uw,Caracciolo:2017xyc}.

Recently, it has been shown that the $M=1$ sigma model has no phase transition in two dimensions, contrary to usual Bernoulli percolation \cite{Bauerschmidt:2019mhu}. (This was done by using supersymmetric localization, in order to first show that, for supersymmetric observables, integrals over $H^{0|2} $ are equivalent to integrals over $H^{2|4} $.)  It was shown later that at and above three dimensions, there is a transition \cite{Bauerschmidt:2021nfm}.

In fact, for $g^2<0$ this sigma model is perturbatively asymptotically free in two dimensions, and so has a UV fixed point in $d=2+\epsilon $ (with $\epsilon >0$) dimensions. In \cite{Klebanov:2021sos}, a UV completion of these theories in $d>2$ was proposed beyond $M=1$ \cite{Fei:2015kta}.
It is a UV completion in the sense that it has a trivial UV fixed point, while in the IR it flows to the fixed point of the sigma model.
The proposed action is given by
\begin{equation} \label{eq:UV_completion_OSp}
\int d^dx \left( \partial _{\mu } \bar \theta ^i \partial ^{\mu } \theta ^i + \frac{1}{2} (\partial _{\mu } \sigma )^2+ g(\sigma ^2 +2 \bar \theta ^i \theta ^i)^{(2M+1)/2} \right)
\end{equation}
having upper critical dimension $d_c=2 \frac{2M+1}{2M-1} $.
The first theory in this family is when $M=1$ with action
\begin{equation} \label{eq:UV_completion_OSp_M_1}
\int d^dx\left( \partial _{\mu } \bar \theta \partial ^{\mu } \theta + \frac{1}{2} (\partial _{\mu } \sigma )^2 + g\sigma ^3 + 3g\sigma \bar \theta \theta \right) .
\end{equation}
This situation is similar to the Gross-Neveu model in $2<d<4$ which is non-renormalizable, but is described by the renormalizable Gross-Neveu-Yukawa model \cite{Hasenfratz:1991it,Zinn-Justin:1991ksq}.

The case $M=2$ is of particular interest as it has an upper critical dimension of $10/3$ which is above three dimensions. Interestingly, the $\epsilon $ expansion for three dimensions is expected to be more accurate than usual since $\epsilon =1/3$ is small. Correspondingly, the action
\begin{equation} \label{eq:UV_completion_OSp_M_2}
\int d^dx \left( \partial _{\mu } \bar \theta ^i \partial ^{\mu } \theta ^i + \frac{1}{2} (\partial _{\mu } \sigma )^2+ \frac{g_1}{6} \sigma (\bar \theta ^i\theta ^i)^2+\frac{g_2}{6} \sigma ^3 \bar \theta ^i \theta ^i+\frac{g_3}{120} \sigma ^5\right) 
\end{equation}
was considered in \cite{Klebanov:2021sos}.
Expressed using real Grassmann coordinates, this theory only has an $Sp(4,\mathbb{R} )$ symmetry.
However, it was shown in \cite{Klebanov:2021sos} that close to the upper critical dimension, in $d=10/3-\epsilon $, the IR fixed point of the theory has couplings that are in proportion so as to give the $OSp(1|4)$ invariant potential (i.e., a power of $\sigma ^2+2\bar \theta ^i \theta ^i$). A similar symmetry enhancement was shown to occur for $M=3$ in \cite{Klebanov:2021sos}.

Here we are interested first in deriving a spanning forest description for the $OSp(1|2M)$ theories. This can be useful in applying Monte Carlo simulations to such random cluster models, and confirming using them the upper critical dimensions (by observing at what dimensions the critical exponents become those of free field theory) as well as finding numerically the critical exponents in different space dimensions. We would also like to extend the correspondence to the Potts model that is known for the $M=1$ sigma model, and was shown to be consistent with the $\epsilon $-expansion of the $M=1$ field theory \eqref{eq:UV_completion_OSp_M_1} in \cite{Fei:2015kta}.
We are particularly interested in $M=2$.
In order to do that, we assume that the sigma model \eqref{eq:OSp_sigma_model} and the field theory \eqref{eq:UV_completion_OSp} have the same critical behavior, so that we can prove our claim using the sigma model. We start by deriving a fermionic representation of the sigma model, for which we know the spanning forest representation, and have derived an equivalent Potts theory.

\subsection{Fermionic and forest representations}

Let us start by solving the constraint \eqref{eq:OSp_sigma_model_constraint} of the sigma model classically. The solution can be written as
\begin{equation}
\sigma  = \sum _{k=0} ^M a_k (\bar \theta ^i \theta ^i)^k = e^{-\sum _{k=1} ^M b_k (\bar \theta ^i \theta ^i)^k} 
\end{equation}
where the sum can be truncated at $M$ since the higher powers vanish by the Grassmann algebra.
The coefficients are given by
\begin{equation}
a_k = (-2)^k \binom{1/2}{k},\qquad b_k=\frac{2^{k-1} }{k} 
\end{equation}
The first few values are $a_k=(1,-1,-\frac{1}{2} ,-\frac{1}{2} ,-\frac{5}{8} ,\cdots ) $ (starting at $k=0$) and $b_k=(1,1,\frac{4}{3} ,2,\cdots )$ (starting at $k=1$).

Thinking for the moment about this as a statistical model at inverse temperature $\beta $ in dimension $d$, the partition function of the model \eqref{eq:OSp_sigma_model} is (below we restrict the delta function to the solution above, and do not include the other solution by definition)
\begin{equation}
\begin{split}
& \int D\sigma D\theta D\bar \theta \,  \delta (\sigma ^2-1+2\bar \theta ^i \theta ^i) \exp \left[ -\frac{\beta }{2g^2} \int d^dx \left( (\partial _{\mu } \sigma )^2+2\partial _{\mu } \bar \theta ^i \partial ^{\mu } \theta ^i\right) \right] .
\end{split}
\end{equation}
Let us discretize the Euclidean space to be a lattice with spacing $a$. Then we get up to an overall factor of a power of 2 (from the delta functions)
\begin{equation}
\begin{split}
&Z = \int D\sigma  D\theta D\bar \theta  \prod _x \frac{1}{\sigma _x} e^{-\beta S} \prod_x \delta \left( \sigma _x - \sum _{k=0} ^M a_k (\bar \theta ^i_x \theta ^i_x)^k \right) = \\
& = \int D\theta D\bar \theta \, \exp \Bigg[ \sum _x \sum _{k=1} ^M b_k (\bar \theta ^i_x \theta ^i_x)^k+ \frac{\beta a^{d-2} }{g^2} (-2d) \sum _x \bar \theta ^i_x \theta ^i_x + \frac{\beta a^{d-2} }{g^2} \sum _{\langle xy\rangle } (\bar \theta ^i_x \theta ^i_y + \bar \theta ^i_y \theta ^i_x)+\\
&\qquad + \at[ \frac{\beta a^{d-2} }{2g^2} (-2d)\sum _x \sigma _x^2 +\frac{\beta a^{d-2} }{g^2} \sum _{\langle xy\rangle } \sigma _x \sigma _y {\Bigg]} ]{\sigma _x=\sum _k a_k (\bar \theta ^i_x \theta ^i_x)^k} .
\end{split}
\end{equation}
For $\sigma ^2$ we can substitute the constraint in its original form
\begin{equation}
\begin{split}
& Z= \int D\theta D\bar \theta \, \exp \Bigg[ \sum _x \sum _{k=1} ^M b_k (\bar \theta ^i_x \theta ^i_x)^k 
+ \frac{\beta a^{d-2} }{g^2} (-2d)\sum _x \bar \theta ^i_x \theta ^i_x + \frac{\beta a^{d-2} }{g^2} \sum _{\langle xy\rangle } (\bar \theta ^i_x \theta ^i_y+\bar \theta ^i_y \theta ^i_x) - \\
& - \frac{\beta a^{d-2} }{g^2} d \sum _x (1-2\bar \theta ^i_x \theta ^i_x) + \frac{\beta a^{d-2} }{g^2} \sum _{\langle xy\rangle } \sum _{k,l=0} ^M a_k a_l (\bar \theta ^i_x \theta ^i_x)^k (\bar \theta ^j_y \theta ^j_y)^l \Bigg] .
\end{split}
\end{equation}
It is useful to explicitly write the first leading terms in the polynomials, resulting in several cancelations
\begin{equation} \label{eq:fermionic_OSp_preliminary_form}
\begin{split}
& Z= \int D\theta D\bar \theta  \, \exp \Bigg[ 
\sum _x \sum _{k=1} ^M b_k (\bar \theta ^i_x \theta ^i_x)^k 
- \frac{\beta a^{d-2} }{g^2} 2d \sum _x \bar \theta ^i_x \theta ^i_x
+ \frac{\beta a^{d-2} }{g^2} \sum _{\langle xy\rangle } (\bar \theta ^i_x \theta ^i_y + \bar \theta ^i_y \theta ^i_x) - \\
& \qquad 
-\frac{\beta a^{d-2} }{g^2} d \sum _x (1)
+ \frac{\beta a^{d-2} }{g^2} \sum _{\langle xy\rangle } \sum _{k+l>1} ^M a_k a_l (\bar \theta ^i_x \theta ^i_x)^k(\bar \theta ^j_y \theta ^j_y)^l 
+ \frac{\beta a^{d-2} }{g^2} \sum_x (d) \Bigg] .
\end{split}
\end{equation}

Let us mention that while in the sigma model everything depends on the bilinear form, and so the $OSp(1|2M)$ symmetry is manifest, this is no longer the case in this fermionic theory.
The induced symmetry transformation is obtained by transforming the fermions only, and substituting for any occurrence of the boson the result of the constraint. This gives a nonlinear transformation. In addition, the naive fermionic measure is not invariant under it. Together with the transformation of the measure, and the action, the full integral is invariant. This was done in \cite{Jacobsen:2005uw,Caracciolo:2007mg}.
Note, however, that this supersymmetry transformation becomes trivial under \eqref{eq:equiv_ferminic_Potts}, which maps only observables with even number of fermionic fields.

\begin{figure}[h]
\centering
\includegraphics[width=0.4\textwidth]{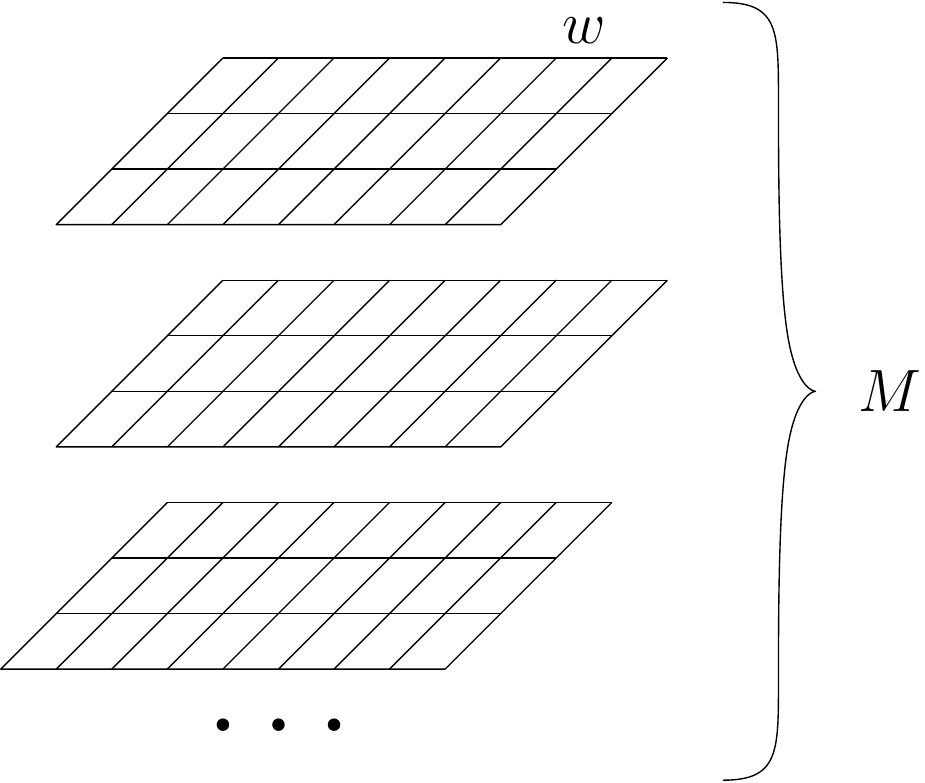}
\caption{The graph corresponding to the sigma model with $M$ complex fermions, demonstrated in two dimensions.}
\label{fig:cubic_lattice_copies}
\end{figure}

This expression \eqref{eq:fermionic_OSp_preliminary_form} is of the form \eqref{eq:fermionic_theory_no_edges} and so can be described using spanning forests. Let us define
\begin{equation}
w = - \frac{\beta a^{d-2} }{g^2} .
\end{equation}
The structure of the graph $G$ is determined by the quadratic terms. More precisely, the off-diagonal quadratic terms determine the graph. To form the Laplacian term in \eqref{eq:fermionic_theory_no_edges}, we should add to them quadratic diagonal terms that are fixed by the off-diagonal ones, so that the Laplacian matrix has vanishing sum of rows and columns, as we reviewed in section \ref{sec:duality}. The remaining quadratic diagonal terms are treated as mass terms so that they are perturbations in the spanning forests language.
In our case, the graph is given by $M$ copies of a simple cubic lattice. There are no links between the $M$ copies. Within each copy, every two nearest neighbor sites are connected by one edge, with all edges being of the same weight $w$. This is shown in Fig.\ \ref{fig:cubic_lattice_copies} for simplicity in two dimensions. Using the Laplacian matrix corresponding to this graph, we can write the partition function as
\begin{equation} \label{eq:forest_form_of_partition_function}
\begin{split}
&Z = \int D\theta D\bar \theta  \, \exp \Bigg[ \bar \theta L\theta + \sum _x \sum _{k=1} ^M b_k (\bar \theta ^i_x \theta ^i_x)^k - w \sum _{\langle xy\rangle } \sum _{\substack{k,l=0 \\ k+l>1}} ^M a_k a_l (\bar \theta ^i_x \theta ^i_x)^k (\bar \theta ^j_y \theta ^j_y)^l \Bigg] .
\end{split}
\end{equation}

\begin{figure}[h]
\centering
\includegraphics[width=0.4\textwidth]{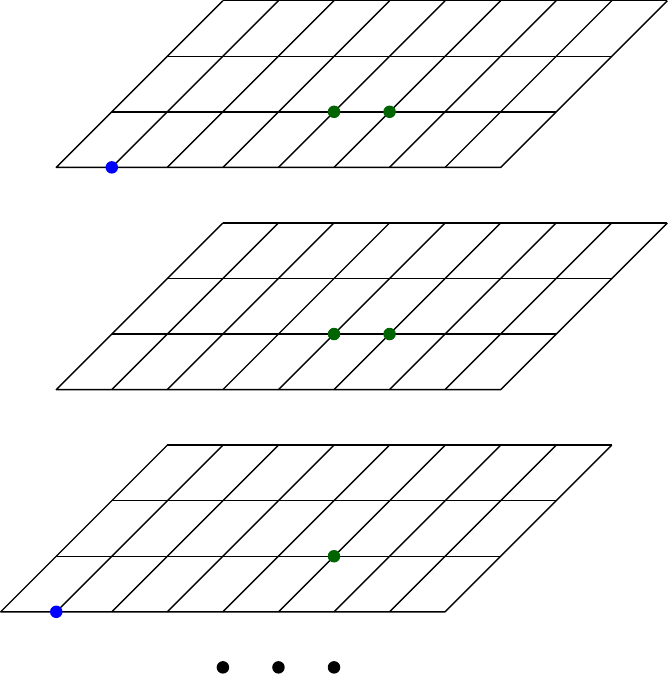}
\caption{The types of subgraphs included in marking the trees of the forests.}
\label{fig:allowed_subgraphs}
\end{figure}

This partition function has the representation \eqref{eq:general_spanning_forests_formula}. We can treat the interactions as containing no edges, and then the subgraphs $H$ in the statistical sum are spanning forests.
In this case there are two kinds of elements of the interaction subgraphs $\Gamma $. One is given by a subset of the same vertex $x$ of the $M$ copies. The other is a subset of a pair of nearest neighbor vertices in the $M$ copies. The two vertices do not have to appear together in each copy. An example is shown in Fig.\ \ref{fig:allowed_subgraphs}.

Therefore, we get that the $OSp(1|2M)$ sigma models are described equivalently by a spanning forests random model \eqref{eq:general_spanning_forests_formula}. In this formula, we should sum over spanning forests $H$.
The subgraphs $\Gamma $ of the interactions used to mark the trees, and the corresponding coefficients $t_{\Gamma } $ are
\begin{itemize}
\item 
The mass term comes from the $b_1$ term. It corresponds to a subgraph $\Gamma $ consisting of a single vertex. The corresponding coefficient $t_{\Gamma } $ is $b_1=1$.
\item
Another class of $\Gamma $'s is the subgraph consisting of $k > 1$ distinct vertices, all in the same coordinates $x$, in $k$ out of the $M$ copies. Such a $\Gamma $ comes with a coefficient $t_{\Gamma } $ given by $b_k k!-2dwa_k k! $.
\item 
The last set of subgraphs consists of a nearest neighbor pair $xy$, where $x$ appears in any number $k$ out of the $M$ copies, and $y$ appears in $l$ out of the $M$ copies (with $k,l \ge 1$), and the corresponding coupling is $(-w) \cdot k! \cdot l! \cdot a_k a_l$.
\end{itemize}

\begin{figure}[h]
\centering
\includegraphics[width=0.8\textwidth]{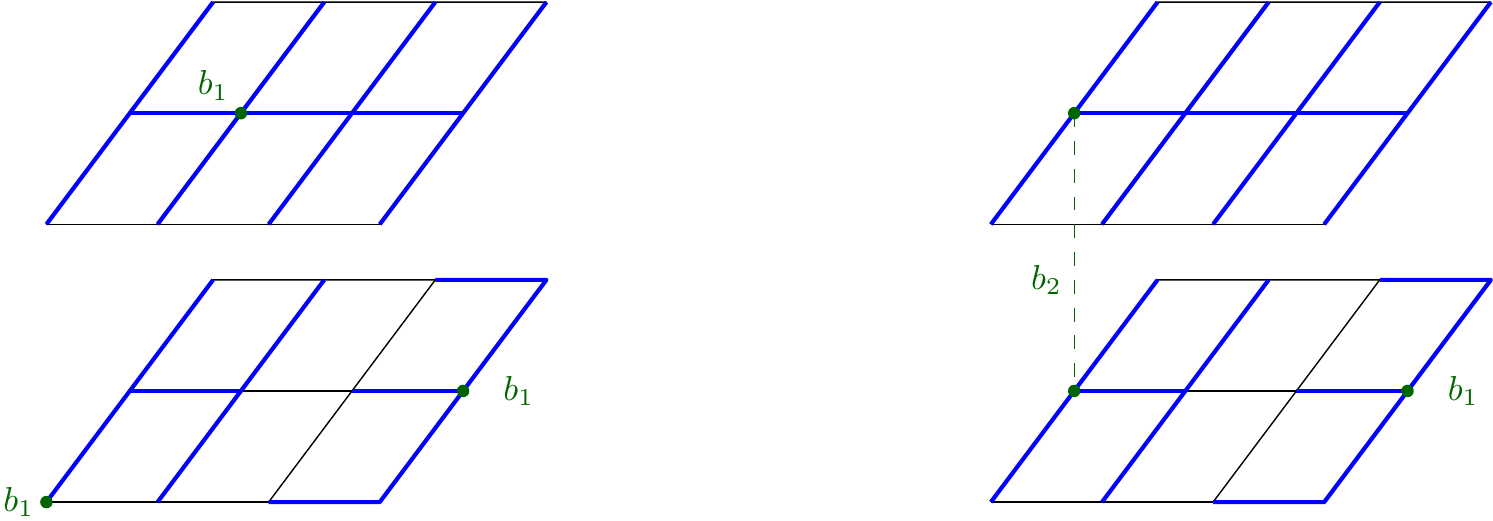}
\caption{Two examples of possible markings for $M=2$. In both cases, we show the same forest $H$ in blue (and using heavier lines, so that the color is not essential), and the roots (or marks) in green, as well as mark them by heavy dots.}
\label{fig:b_markings}
\end{figure}

In order to demonstrate the forest description in our case, let us consider for simplicity the case of $M=2$ in two dimensions and include for demonstration just the $b_k$ interactions in \eqref{eq:forest_form_of_partition_function}. The interactions are
\begin{equation}
\sum _x b_1 \bar \theta ^1_x \theta ^1_x + \sum _x b_1 \bar \theta ^2_x \theta ^2_x + 2b_2 \sum _x \bar \theta ^1_x \theta ^1_x \bar \theta ^2_x \theta ^2_x .
\end{equation}
The partition function is given by a sum over spanning forest subgraphs $H$ of the graph $G$, where $G$ is two copies of the square lattice. For each $H$ first we give the usual weight of the product of edges. In addition, we should go over all the possibilities of marking it with the couplings from the interactions. An example of a particular $H$ is shown in Fig.\ \ref{fig:b_markings}. When we use the $b_1$ interactions which are connected, each connected component of $H$ (a tree) should have exactly one $b_1$ marking. The positions of the markings can be anything inside a component, and independent among the different components. So this contribution simply factorizes, and is the same as what we would have obtained in the product of partition functions of $M=1$ if we had only the $b_1$ interaction. This is expected since in this case the interaction in the Hamiltonian factorizes. However, the $b_2$ interaction marks two points on the two copies of the lattice, with the points having the same coordinates. A single $b_2$ interaction therefore marks two connected components. This gives a correlation between the two copies of the lattice, and the result no longer factorizes.

\subsection{Dual Potts model for the $OSp(1|4)$ theory} \label{sec:M_2_Potts}

Given the fermionic representation \eqref{eq:forest_form_of_partition_function}, we can now apply the result of section \ref{sec:duality} in order to get a Potts dual to the $OSp(1|2M)$ theories.
For these theories, the corresponding graph consists of $M$ copies of the lattice, as we saw. Therefore we should use Potts variables $\sigma _{x,i} $ with $i=1,\cdots ,M$.

Let us consider specifically the case of $M=2$. We listed the interactions, that is the set of $\Gamma $'s before, in \eqref{eq:forest_form_of_partition_function}. Using \eqref{eq:equiv_ferminic_Potts_Hamiltonian}, we find the Hamiltonian (with the first term being fixed by the graph structure)
\begin{equation}
\begin{split}
H &= - \sum _{\langle xy\rangle } \sum _{i=1} ^2 \log(1+\sqrt{q}w) \delta _{\sigma _{x,i} ,\sigma _{y,i} } -\sum _{x} \sum _i \log(1+\sqrt{q})\delta _{\sigma _{x,i} ,h} -\\
& -\sum _x \log\left( 1+q(2+2dw)\right) \delta _{\sigma _{x,1} ,h} \delta _{\sigma _{x,2} ,h} -\sum _{\langle xy\rangle } \sum _{ij} \log(1-wq)\delta _{\sigma _{x,i} ,h} \delta _{\sigma _{y,j} ,h} -\\
& - \sum _{\langle xy\rangle } \sum _i \log(1-wq^{3/2} ) \left( \delta _{\sigma _{x,i} ,h} \delta _{\sigma _{y,1} ,h} \delta _{\sigma _{y,2} ,h} +(x \leftrightarrow y)\right) - \\
& - \sum _{\langle xy\rangle } \log(1-wq^2)\delta _{\sigma _{x,1} ,h} \delta _{\sigma _{x,2} ,h} \delta _{\sigma _{y,1} ,h} \delta _{\sigma _{y,2} ,h} .
\end{split}
\end{equation}

We have proven that this theory is dual to the $M=2$ fermionic theory we constructed, which in turn we showed to be equivalent to the sigma model for $M=2$. We propose that it also has the same critical behavior as the $OSp(1|2M)$ field theory for $M=2$, Eq.\ \eqref{eq:UV_completion_OSp_M_2}.

Note that in order to do perturbation theory in $g^2$, it is more natural to rescale $\theta \to w^{-1/2} \theta $ and similarly for $\bar \theta $. Doing that, we get instead the Potts model
\begin{equation}
\begin{split}
H_{\text{pert}}  &= - \sum _{\langle xy\rangle } \sum _{i=1} ^2 \log(1+\sqrt{q}) \delta _{\sigma _{x,i} ,\sigma _{y,i} } -\sum _{x} \sum _i \log(1+\sqrt{q}w^{-1} )\delta _{\sigma _{x,i} ,h} -\\
& -\sum _x \log\left( 1+q(2w^{-2} +2dw^{-1} )\right) \delta _{\sigma _{x,1} ,h} \delta _{\sigma _{x,2} ,h} -\sum _{\langle xy\rangle } \sum _{ij} \log(1-w^{-1} q)\delta _{\sigma _{x,i} ,h} \delta _{\sigma _{y,j} ,h} -\\
& - \sum _{\langle xy\rangle } \sum _i \log(1-w^{-2} q^{3/2} ) \left( \delta _{\sigma _{x,i} ,h} \delta _{\sigma _{y,1} ,h} \delta _{\sigma _{y,2} ,h} +(x \leftrightarrow y)\right) - \\
& - \sum _{\langle xy\rangle } \log(1-w^{-3} q^2)\delta _{\sigma _{x,1} ,h} \delta _{\sigma _{x,2} ,h} \delta _{\sigma _{y,1} ,h} \delta _{\sigma _{y,2} ,h} .
\end{split}
\end{equation}
Indeed, the interaction terms are suppressed by at least $g^2$. This includes the ``mass term'' in the fermionic description (the interaction quadratic in the fermions, corresponding to marking with a single vertex in the spanning forest description), which becomes a magnetic field term in the Potts language.

\section{Forest representations of the $OSp(1|4)$ theory} \label{sec:M_2_forest}

In section \ref{sec:fermionic_OSp} we described explicitly the spanning forest description of the $M=2$ theory, but we included only part of the interactions for demonstration purposes. In this section, we give the full random spanning forest model. In addition, when we consider disconnected interactions as we did above, we have contributions where we mark several connected components by a single interaction. This is similar in a sense to a non-local interaction. One may wish to have a description where this does not happen. In particular, this is expected to be more useful when simulating such statistical models. We will give in this section an equivalent spanning forest model where the components $H_i$ in \eqref{eq:general_spanning_forests_formula} are always connected.

The full $M=2$ theory, where now we include all the interactions, has a partition function given by a sum over spanning forests as in equation \eqref{eq:general_spanning_forests_formula}. Each edge of the forest is assigned the value $w$. The forests should be fully marked, where the markings are a special case of what we found in section \ref{sec:fermionic_OSp}, and are concretely:
\begin{itemize}
\item 
A single point in either of the two lattices, assigned the value $1$.
\item 
The point with the same coordinates marked on both lattices, assigned the value $2+2dw$.
\item 
Two nearest neighbor points, each one appearing in one of the two lattices (could be the same one), assigned the value $(-w)$.
\item 
Two nearest neighbor points, one appearing once and the other appearing in both lattices, assigned the value $(-w)$.
\item 
Two nearest neighbor points appearing in both lattices, with the value $(-w)$.
\end{itemize}

Given the description above, let us provide one direction to simplify the forest description of such a theory. The idea is based on the comment mentioned in section \ref{sec:duality} that relates a disconnected nearest neighbor marking and the equivalent connected marking that differs only by the coupling constant. The main idea will be that we can trade a forest with a particular marking with another forest and its marking, in case there is a good correspondence between the two.

Let us start with two markings mentioned above, namely the one assigning to every vertex the value $1$ and the one assigning to a pair of nearest neighbors on the same lattice the value $(-w)$. Let us repeat and rephrase the trading of the disconnected latter marking with a connected one. The disconnected $(-w)$ term can mark an $H_i$ if it has two tree components, each one having a vertex from the neighboring pairs. We can trade such a marking in favor of considering the tree obtained by adding the edge connecting the two neighbors to $H_i$, and marking this edge (instead of the two vertices on its ends). This gives us a good correspondence, since connecting two disconnected trees by a path is in one-to-one correspondence with the set of trees containing the added path (indeed, we cannot have a cycle since this would contradict the fact that the trees were disconnected before). This step is illustrated in Fig.\ \ref{fig:trade_nn_marking}.

The combination of this marking with the one of vertices is easy to describe. It gives for each vertex of a tree a value of $1$ and for each edge a value of $(-1)$ (since the $w$ is absorbed into the edge value). They combine to give just $1$ in total for every tree. In the case of $M=1$, these are the only markings that we have, and so the partition function is just
\begin{equation}
\sum _{\substack{\text{Spanning}\\ \text{forests } F}} \prod _{e \in F} w_e .
\end{equation}
This is the same as the $q \to 0$ limit of the Potts model with nearest neighbor interaction. Note that in this description, the forests can be considered as unrooted, since we do not need to sum over forests together with their possible roots.

We can now extend the idea above. For the marking of two points with the same coordinates on both lattices, we can trade it with a marking where again we connect the two points by an edge, and mark this edge, giving the value $2+2dw$. Note that we assign $w_e=1$ for edges going between the lattices. In fact, there are no such edges in the original construction. This is important, since such ``virtual forests'' differ from the original forests that we had. In the latter, we had to sum over all the markings of such forests. However, in the newly constructed forests, we do not build interactions on top of them, but rather \emph{only} include the markings that were constructed by such a trade of markings. So, now we do have forests that connect the two lattices, but only with the markings that we specify.

\begin{figure}[h]
     \centering
     \begin{subfigure}[b]{0.3\textwidth}
         \centering
         \includegraphics[width=\textwidth]{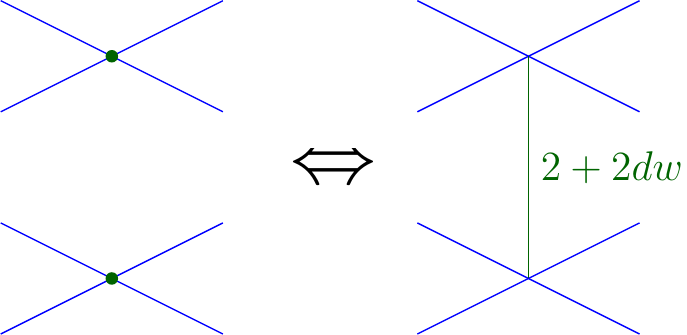}
         \caption{}
     \end{subfigure}
     \qquad\qquad\quad\,\,\,\,\,\,\,
     \begin{subfigure}[b]{0.3\textwidth}
         \centering
         \includegraphics[width=\textwidth]{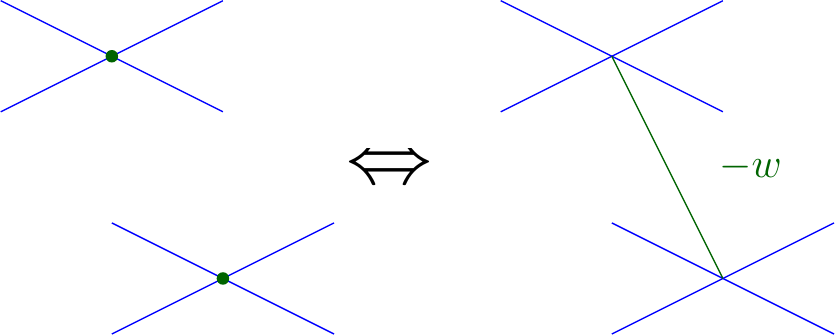}
         \caption{}
     \end{subfigure}
     \hfill
     \par \bigskip \bigskip
     \begin{subfigure}[b]{0.3\textwidth}
         \centering
         \includegraphics[width=\textwidth]{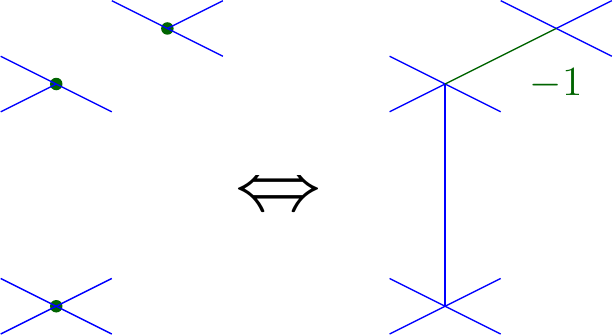}
         \caption{}
     \end{subfigure}
     \qquad\qquad\quad\,\,\,\,\,\,\,
     \begin{subfigure}[b]{0.3\textwidth}
         \centering
         \includegraphics[width=\textwidth]{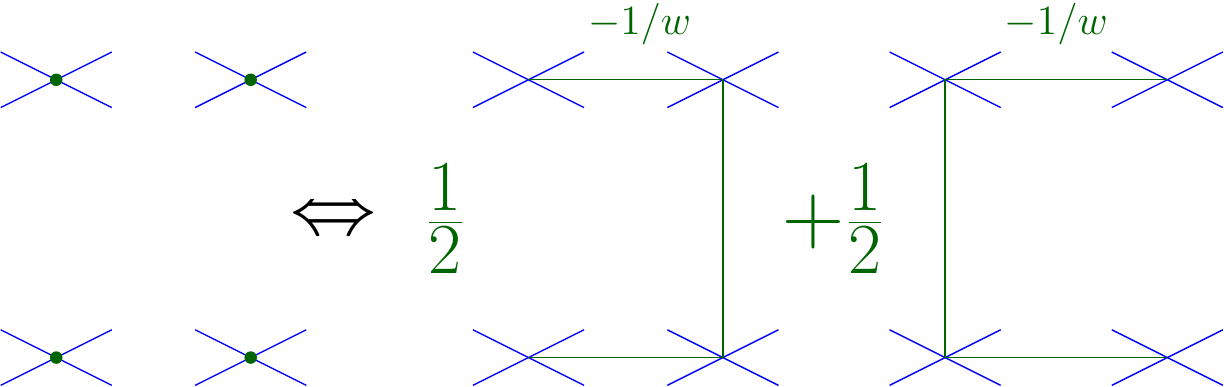}
         \caption{}
     \end{subfigure}
     \hfill
    \caption{New markings in $M=2$.}
    \label{fig:trading_new_markings}
\end{figure}


Similarly, we can do with the two nearest neighbor pairs that appear each in a different lattice, giving us forests with edges that go between two neighbors but each one appearing on a different lattice. The weight of this marking is $(-w)$ as the edge between the lattices is assigned the value $1$. These are shown in Fig.\ \ref{fig:trading_new_markings}. For the case of two neighbors where one appears in one lattice and the other appears in both, we can connect now both the two points with the same coordinates between the lattices, and the two neighbors on the same lattice. This gives us the same kind of tree as in the first case, with a vertical edge going between the lattices, but now there is a new marking of an edge inside a lattice that is attached to the vertex with the inter-lattice connection. The value assigned is $(-1)$ (since again $w$ becomes an edge factor). Lastly, for two neighbors appearing in both lattices, we can introduce yet another marking of the same trees with vertical inter-lattice connections. This is a marking of an edge that appears in both lattices and connects to the inter-lattice vertices. The weight is $\left( -\frac{1}{2w}\right) $, see Fig.\ \ref{fig:trading_new_markings}.

With these manipulations, we now have only connected markings, and so we can restrict the formula to $H_i$'s that are connected. The price to pay is that the weights assigned to trees are different depending on the type of tree. Concretely, we have
\begin{equation} \label{eq:M_2_forest_rep}
\begin{split}
Z_{ M=2}   = \sum _{\substack{H=(H_1,\cdots ,H_l) \text{ spanning} \\ H_i \text{ trees}}} \left( \prod _{i=1} ^l W(H_i)\right)  \prod _{e \in H} w_e
\end{split}
\end{equation}
where now there are also edges connecting the two lattices (vertical ones, and edges going between nearest neighbors in terms of their coordinates).
$W(H_i)$ is given by
\begin{itemize}
\item 
For $H_i$ a tree embedded in one lattice, $W(H_i)=1$.
\item 
For $H_i$ having a non-vertical (but nearest neighbor) inter-lattice connection, $W(H_i)=-w$ as there is a single marking of the inter-lattice edge.
\item 
For a tree with a vertical edge connecting the two lattices, with endpoints $v_1$ and $v_2$, we have $W(H_i) = 2+2dw-d(v_1)-d(v_2) - \frac{1}{2w} d(v_1 \cap v_2)$ where $d(v_i)$ is the degree of $v_i$ in $H$ in its lattice, and $d(v_1 \cap v_2)$ is the number of edges with the same coordinates appearing in both lattices and belonging to $H$ (one connecting to $v_1$ and the other to $v_2$).
\end{itemize}

This gives a forest representation of this problem where we have only connected components. However, note that while forests can now go between the two lattices, there can be such an inter-lattice connection only up to at most once in each tree. We should mention that this is a sort of a non-local constraint on the trees.

\section{Concluding remarks}

We have discussed a dual Potts model for fermionic theories of the form \eqref{eq:fermionic_theory_no_edges} by using a spanning forest description. A simple way to phrase this description is that we have a random model of forests, that is, a gas of trees, each one assigned the value of the edges. For each forest, we need to go over all the ways to mark it with coupling constants, such that each tree has one root which is part of the interactions. Altogether we have random rooted forests. The trees correspond to regions where the Potts degrees of freedom align in the same direction.

We have used the fact that we can absorb all the edges in the subgraphs defining the interactions into the coupling constants, and thus could consider interactions made of vertices only. It would be interesting to provide a Potts description for these fermionic theories when we explicitly include edges as well. While the result is the same, different ways to describe it could be useful in the future. For example, marking with connected components may be simpler to simulate. One approach to achieve this while still using the form \eqref{eq:fermionic_theory_no_edges} was described in section \ref{sec:M_2_forest}.

Here we consider a fixed lattice for the theory, while critical phenomena on a random graph, which can be thought of as gravity, has been studied in \cite{Kazakov:1987qg, Caracciolo:2009ac, Bondesan:2016osa, Kazakov:1988fv, Daul:1994qy}.
This results in a random matrix theory, where the Potts degrees of freedom become matrices. It would be interesting to study the generalized matrix theory corresponding to the theories we introduce here \cite{KazakovInProgresss}. One possible approach is to group several interacting vertices in the matrix theory and represent them by an effective interaction vertex, with a coefficient having a  higher power of the vertex counting parameter (for context, see \cite{Kazakov:1987qg}).

There has been several recent developments in understanding the $OSp(1|2M)$ theories, as we mentioned. In particular, for $M=1$, it is well established that there is no phase transition in two dimensions, while there is a transition in higher dimensions \cite{Bauerschmidt:2019mhu,Bauerschmidt:2021nfm}. It is still important to understand what is the upper critical dimension for second order transitions. The recently introduced \cite{Klebanov:2021sos} field theories \eqref{eq:UV_completion_OSp} provide a suggestion for the upper critical dimension for any $M \ge 1$, and it would be interesting to test this. The spanning forests discussed here or the Potts model could be useful for that.

\appendix

\section*{Acknowledgments}

I would like to thank Roland Bauerschmidt, Jesper Jacobsen, Vladimir Kazakov, Hubert Saleur, and Bernardo Zan for useful discussions. I am especially thankful to Igor Klebanov for many discussions which prompted this work.

\section{Continuation of vector models under $N \to -N$} \label{sec:NToMinusN}

As was mentioned in the main text, the continuation of a bosonic vector model under $N \to -N$ is given by the same vector model where the commuting variables are replaced by anticommuting ones. The familiar argument for this is that every anticommuting loop gives a factor of $N$ and comes with a minus sign. There are two drawbacks to this argument. First, it is manifestly valid only when the action is first order in the bilinear anticommuting term (that is, in interactions of the Yukawa type). This is why it is customary to introduce a Hubbard–Stratonovich field in other cases, such as in the quartic vector model. In the cubic model, given by \eqref{eq:UV_completion_OSp} with $M=1$, the action is already first order in the bilinear term. Second, this argument is perturbative in nature.

Here we give a short and formal argument for the same claim, which is not perturbative in nature, at least for a regularized theory. Consider a vector model with variables $\psi^i$. We can write the partition function as
\begin{equation}
\begin{split}
& \int D\psi^i \exp\left[ -\int dx \left( \partial _{\mu } \psi^i \partial ^{\mu } \psi^i+f(\bar \psi^i \psi^i)\right) \right]  = \\
& = \int DG D\Sigma D\psi^i \exp \Bigg[ i \int dxdy \, \Sigma (x,y)\left( G(x,y)-\sum _i \bar \psi^i(x)\psi^i(x)\right)  -\\
& \qquad \qquad - \int dx \left( \pder{}{x^{\mu } } \pder{}{y_{\mu } } G(x,y) \at{x=y} + f\left( G(x,x)\right) \right) \Bigg] = \\
& = \int DG D\Sigma \,  e^{ i \int dx dy \, \Sigma (x,y)G(x,y) - \int dx \left( \pder{}{x^{\mu } } \pder{}{y_{\mu } } G(x,y) \at{x=y} + f\left( G(x,x)\right) \right) } \left[ \det \Sigma (x,y)\right] ^{\pm N} 
\end{split}
\end{equation}
where the plus sign corresponds to anticommuting fields and the minus sign to commuting fields. We see that the only dependence on the statistics of the fields and on $N$ comes in this exponent. Therefore, changing the statistics of the fields is accomplished by sending $N \to -N$.

\bibliography{main}
\bibliographystyle{JHEP}
\end{document}